# Application of Blockchain in Booking and Registration Systems of Securities Exchanges

Mahdi H. Miraz
Centre for Financial Regulation and Economic Development (CFRED)
The Chinese University of Hong Kong (CUHK)
Sha Tin, Hong Kong
m.miraz@cuhk.edu.hk

David C. Donald
Centre for Financial Regulation and Economic Development (CFRED)
The Chinese University of Hong Kong (CUHK)
Sha Tin, Hong Kong
dcdonald@cuhk.edu.hk

*Abstract*—Securities exchange being digitalised and online, security of information and data has become a major concern. Blockchain (BC) technology, being distributed and immutable in nature, has proved to the "Trust Machine" eliminating the need for third-parties. Authors of this paper investigate how Blockchain can be used to secure stock exchange transactions, with an especial focus to the technological as well as legal aspects of such applications. Considering the intricate operational structure of the securities exchange, the research proposes to design, develop and implement a hybrid BC, customised according to the need of the respective stock exchange. The study suggests that the use of such BC can bring many benefits which the other technologies currently being used cannot offer. However, during the design process of any such application using BC, the relevant laws and regulations of the corresponding country need to be considered.

*Keywords—Securities Exchange, Stock Exchange, Blockchain, Distributed Ledger, FinTech, RegTech, LawTech*

## I. INTRODUCTION

A securities exchange can be defined in many different ways and from various perspectives [1]. However, as the name implies it is fundamentally a place (physical or virtual) arranged to facilitate trading in securities like shares of stock and bonds. Stock Exchanges have been shown to play a widely varied but highly significant rôle in the growth and progress of the economy of any country.

To facilitate transacting – which includes both entering into contracts and transferring ownership of securities and cash – stock exchanges have been highly dependent on their infrastructure, which in the last half-century has included technologies for data transfer. Artificial Intelligence, more precisely deep learning, used in algorithmic or "robotic" trading, has added a new dimension. The application of new information technologies to finance, now summed up with the term "FinTech", has always played an important role in the stock exchange. Because stock exchange orders and ownership-related data are transmitted electronically, and ownership is also evidenced electronically (whether ultimately based on certificates or not) security of this data is a major concern. "Bit Commitment" based "Sealed Envelope" [2] and multifaceted applications of Blockchain [3,4,5] including crypto-currencies [6,7] and smart contracts [8,9] possess huge potentials in this regard.

Blockchain, which is in fact a by-product of the Bitcoin [10] crypto-currency system, is a type of Distributed Ledger Technology (DLT). However, from "The Blockchain" of Bitcoin, different variants of Blockchain have evolved. Blockchain technology is increasingly being researched as well as applied in many other domains including the FinTech, RegTech (Regulatory Technology) and LegalTech (Legal Technology) for escalated security and privacy for sensitive data while safeguarding the anonymity of the users [11,12].

A Proof-of-Work (PoW) mathematical puzzle, analogous to Adam Back's HashCash [13], hardens BC security through conserving the digital ledger of transactions from any sort of alternation. To ensure anonymity and safe-guard users' identity, BC utilises changeable Public Key (PK) which provides an extra layer of privacy. Thus BC is basically a group of technologies [14]:

- Cryptographic Algorithms,
- Distributed Network and
- Programme i.e. BC Protocol.

Can Blockchain technology add value to the clearing and settlement systems of securities exchanges? A response to this question requires an examination of both the theoretical potential of blockchain technology and its actual capabilities in the near term. The first question demands explanation of three facets: (i) how democratic control of the ledgers for booking actually works, (ii) how the internal architecture of the ledger prevents alteration, and (iii) what cryptography actually is and why it can stop alteration. This paper in the following sections will address each question in turn.

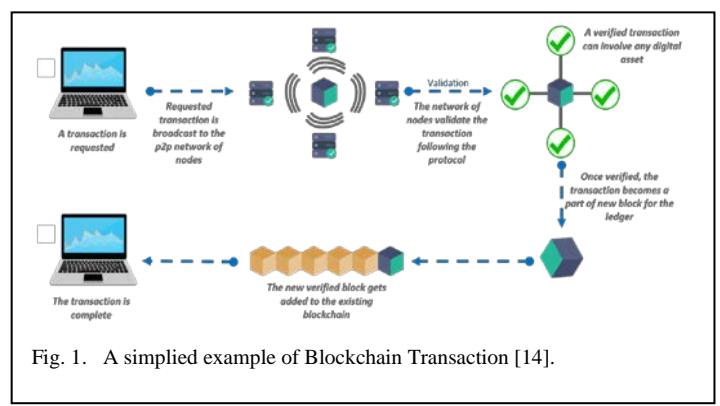

Fig. 1. A simplied example of Blockchain Transaction [14].

## II. BLOCKCHAIN: THE WORLD WIDE LEDGER

Although Bitcon's Blockchain as proposed and later implemented by Satoshi Nakamoto was the first kind of blockchain echo-system [10], it is no longer the only kind. This section presents an anatomy of Bitcoin's BC while highlighting the major differences in other variants.

In its simplest form, a Blockchain is a chain of blocks similar to a series of linked metal rings used for fastening or securing something, or for pulling loads. In a Blockchain, the blocks replace the metal rings of a traditional chain. Each block contains a collection of transactions completed within a certain period of time. To simplify, as shown in Fig 1, once a transaction is triggered by broadcasting it to the BC peer-to-peer (p2p) network of nodes, it is validated by other nodes. The verified transaction is then added to form a part of a new block. Upon completion of the PoW puzzle, the new block is then added to the existing chain and the thus the transaction is complete. Once a new block is added, it is then propagated to the other nodes to verify and append to their exiting copy of BC. Thus, each node has its own copy of the valid and updated BC. Because the BC database is thus replicated among the participating nodes over a distributed network, analogous to the Internet, it is considered to be the World Wide Ledger [9].

To understand this process, we need to understand the inside components of a block. The genesis block, the first block in the chain, is quite different than the other blocks of the chain. It contains smart contracts specifying the rules and regulations to be followed by the nodes for verification and validation purposes as well as for the routine operations of the BC ecosystem. It may optionally contain transactions, especially if any coin is "mined" during the initialisation process of the BC ecosystem.

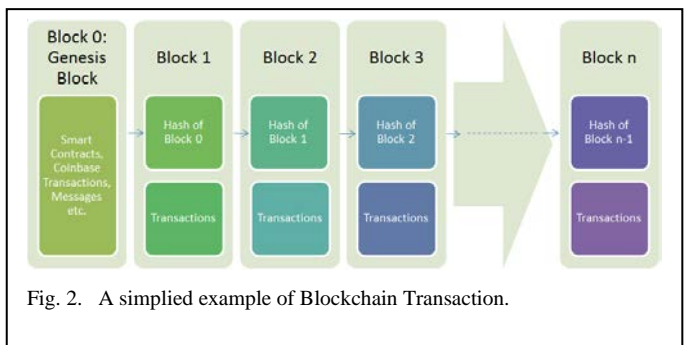

Fig. 2. A simplied example of Blockchain Transaction.

To give it a structure that prevents manipulation, Bitcoin BC possesses dual aims: to let everyone have write access while having no centralised control. Thus the system has been built on a very complex foundation.

In Bitcoin BC, the process of building a block by completing the PoW puzzle is known as mining. "Completing" a PoW puzzle essentially entails calculating the hash of the block that meets certain criteria. Miners spend computational power to solve the PoW, the difficulty level of which is based on how quickly past blocks were solved within a certain timeframe, approximately 10 minutes. As shown in figure-2, each block contains the hash of the previous block. The block header also contains the hash of the current block, calculation of which serves as the backbone of the PoW puzzle. The PoW requires the miner to generate a hash, combining a "nonce" with other data, which is smaller than a certain value; in other words the hash must start with certain numbers of leading zeros.

As shown in figure 3, a block contains version number (4 bytes), hash of the previous block (256 bytes), timestamp [15] (seconds, 4 bytes), Nonce (4 bytes), Bits as the current difficulty level (4 bytes) and Markel [16] root hash the transactions [9,10].

In Bitcoin BC, anyone having a computing device can participate in the BC ecosystem by joining as a node or as part of mining pools which form the nodes. Furthermore, in a programmable BC, smart contract powered machines can also participate as a node. Each node has its own copy of the entire BC public ledger alike a local database. In cooperation with other nodes of the p2p network, each node contributes to conserve the consistency of the chain by making it an immutable one. Furthermore, thus the nodes contribute to make the BC ledger fault tolerance by eliminating Single Point of Failure (SPF).

A transaction can be triggered by any participating node having a private/public pair of cryptographic keys. The triggering node "digitally" signs the transaction by encrypting it using the private key which is then verified by the other nodes by decrypting the transaction message using the corresponding public key. Each public key will only correspond to one private key, and the public key will be broadcasted through the network. Such asymmetric cryptographic authentication system not only provides integrity and non-repudiation but also provides users' identity abstraction across the network. However, asymmetric key cryptography remains secure only as long as the nodes properly manage their respective private keys. The communications amongst the nodes of a BC ecosystem is governed by the corresponding network protocols.

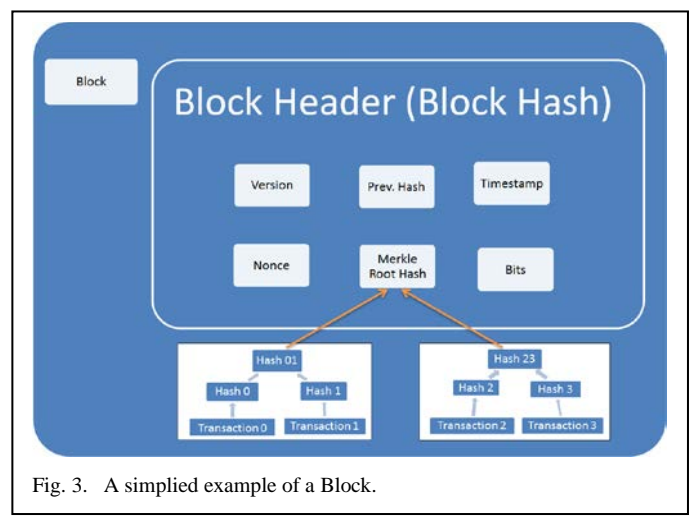

Fig. 3. A simplied example of a Block.

A cryptographic hash function is based on mathematical algorithm that takes any size of data as input and provides a fixed size of output known as the "hash". Hash calculation is very simple and quick; however, tracing back the input data is impossible even if the algorithm is known. Even a single bit change in the input data will produce a completely different hash (non-reversible). A good hash function ensures that the probability of collision is near zero, i.e. two different input data should not generate exactly the same hash. Bitcoin BC uses a very well adopted hashing technique called SHA-256 of the SHA-2 family whereas Ethereum BC uses Keccak-256, both of which produce digests (hash value) of

256 bits. After receiving any transaction data or block, the receiver calculates the hash using the same hashing function and discard/reject the transaction/block if the calculated hash does not match with the included hash. Application of hashing techniques thus helps easily identify any falsification or alternation in a block, thus making the Blockchain secure, tamper-free and virtually un-breakable. Thus BC eliminates the need any third-party such as bank for trust issues, it itself functions as a Trust Machine [17,3] ensuring the required level of trust amongst the participating entities of any transaction.

In fact, not only as part of the PoW puzzle, BC extensively uses hash functions as identifiers of addresses (analogous to a bank account number) as well as transactions as shown in figure 3. However, with the advent of the Merkle Tree [16], all the individual hashes of all the transactions are not needed to be preserved, rather the root has is sufficient. In fact, block in a BC ecosystem is identified by its unique hash values since hash values are collusion free.

Since each block contains the hash of the previous block, it is also not possible to alter even a single transaction data of any past blocks. BC thus establish immutability with the help of PoW consensus approach which requires considerable amount of computing power and electricity to solve the hashing problem with given level of difficulty. Once a transaction is accepted upon verification and validation, it is considered as a raw transaction and piled up in a candidate block to be added to the existing chain. In the meanwhile, the miner keeps trying to solve the PoW puzzle by adding the nonce in calculating the hash value of the block header that matches with the current target value, as explained before. If it doesn't match, the nonce is then modified (usually by adding 1 to it) and the process is repeated until a match is found. The miner, who can solve the PoW puzzle first, completes the block and broadcast it to the other nodes of the network. The other nodes then verify and validate the proposed block. If it is accepted, the other nodes add it to their existing chain and start working on a new block. All the nodes thus have an updated consistent copy of the blockchain. Thus the BC remains secure as long as the share of the computation power of the honest nodes remains higher than the share of dishonest nodes. However, even if a dishonest node can solve the PoW with altered data, it will not be accepted by the other nodes since it will not match with the raw transactions they have in their record.

Since BC maintains a chronological approach in creating and adding blocks to the existing chains along with recording the hash of the previous block in the current block. The hash of the current block thus serves as a cumulative hash of all the previous blocks. As a result, changes in one block will require changes in all the subsequent blocks i.e. to rebuild all the subsequent blocks by successfully solving of the PoW puzzle which will require enormous amount of computational power. To better explain, how this approach of BC makes it immutable, let us consider the following scenarios:

*A. The Double Spending Problem*

For example, Trudy is a dishonest node and would misuse the system by Double Spending. Trudy generates two different transactions at the same time using the same bitcoins: one to another account of herself or her trusted partner and the other one to a retailer. Trudy then broadcasts the payment to the retailer while keeping the other one secret. Once the payment to the retailer gets accepted and added to an "honest" block, the retailer dispatches the product. In the meanwhile, Trudy secretly keeps working on creating a longer block replacing the payment to the retailer by the other one. Once Trudy publishes the secretly built longer chain, the honest nodes are fooled to accept this block based on the "Longest Chain Rule" and keep building on this block. The "honest" block is thus discarded as an "orphan" and the payment to the retailer will be rejected as valid transaction since the bitcoins have already been spent. This is known as the Double Spending Problem. The PoW puzzle makes it harder to avoid creation of such "dishonest" blocks and thus eliminates the possibility of Double Spending. To create the new block, Trudy also has to solve the PoW to create the block hash with the difficulty level set at that time. However, the "honest" miners will also keep working in the meanwhile and may create more honest blocks. Trudy's block has to be longer than the existing block to get accepted. Thus Trudy has to solve the PoW faster than every other node, which is not guaranteed. Furthermore, for this purpose, Trudy has to spend on purchasing powerful devices. It thus becomes extremely hard to take advantage of Double Spending.

The PoW and the Longest Chain Rule thus also prevents any possibility of changing existing data in a Blockchain in the mid-flight i.e. a BC network Trudy is participating in.

*B. Deluding an Audit Team*

Let us consider the scenario where Trudy would like to delude an audit team with a forged off-line copy of the chain. Trudy would like to hide some transactions or add some extra transactions. If the aim is to hide any transaction, the block containing that particular transaction needs to be altered first. If new forged transactions are to be added, since the blocks are timestamped and added chronologically, she cannot just alter the last block. She needs to match the time and date of the transaction and find an appropriate block. For instance, the chain has got 100 blocks and the block needs to be altered is 45. If Trudy makes the intended changes two things will happen: 1) the hash of that particular block (#45) will become invalid and 2) thus the hashes of the subsequent blocks (#46-100) will also become invalid causing the block to fail. To verify chain, the audit team will only require re-calculating the hash of the last few blocks. Trudy can still try to delude the audit team by rebuilding the whole block by replacing hashes of all the subsequent blocks with newly calculated hashes. However, in Bitcoin or any similar BC ecosystems, PoW puzzles needs to be solved for all the blocks which makes the process computationally very expensive. In Multichain or any similar BC ecosystem, the authority to add blocks is by turn and is determined by randomised round robin fashion. The block adders digitally sign the respective blocks. Rebuilding the blocks will require knowing the private keys of all the adders. Thus, in both cases, it is going to be extremely difficult to re-build the whole block.

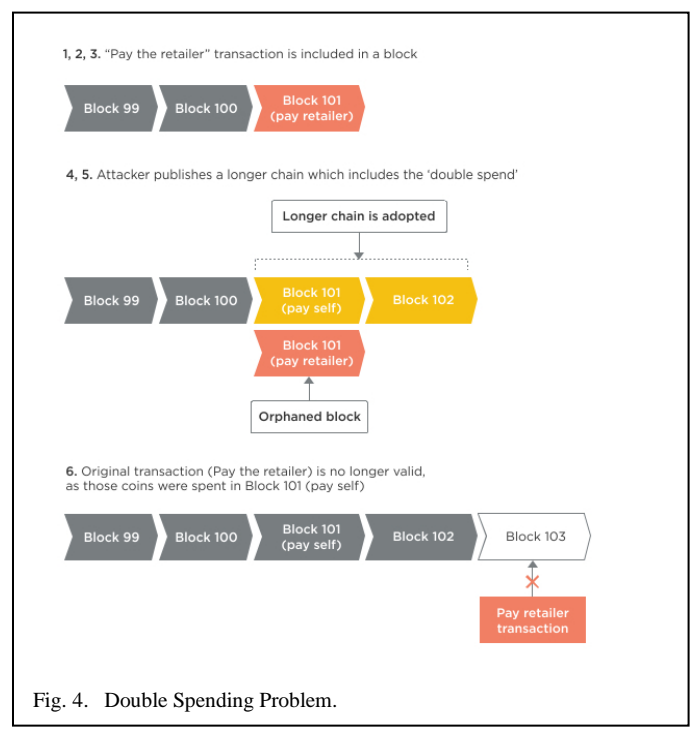

Fig. 4. Double Spending Problem.

For the sake of debate, let us consider that Trudy successfully re-built the block with hiding some transactions or adding some forged ones. However, it is still not possible to delude the audit team. They can simply match the hash of any recent block from the BC supplied by Trudy as well as the hash of the same block from any other (non-colluding) participant. If these two hashes are not matching, even without looking at the data, the audit team will be able to identify that the chain has been altered.

Based on who has the right to write and possesses read permission, Blockchain technology has the following three variants:

### a) Public (Permissionless) Blockchain

Public or Permissionless Blockchain, as the names imply, is open for anyone to join the Blockchain eco-system without requiring any preapproval. Each participant is called a node that has both the right and read access. Examples of such Blockchain ecosystem include: Ethereum, Factom, Blockstream, Bitcoin etc.

### b) Private (Permissioned) Blockchain

On the contrary, Private or Permissioned Blockchain allows only some "trusted" nodes to join the ecosystem who usually has the permission to read and write. However, write permission can also be granted to specific nodes or defined by roles. Example of private Blockchain ecosystem include: Eris Industries, Blockstack, Multichain, Chain and so forth. In fact, in 2015 Chain sealed a partnership deed with NASDAQ to effectuate the application of blockchain ecosystem to enable issuance as well as transfer of shares of privately-held companies securely [18,19]. Chain signed up to be the first company to for rolling out this technology to issue and transfers their shares using NASDAQ's private ledger platform known as NASDAQ Private Market (NPM). However, private BC contradicts with the major motivation of introducing the Bitcoin's BC i.e. to eliminate the need for trusted third-party by establishing complete trust amongst two entities, involved in the transaction, directly [10].

### c) Hybrid (Consortium) Blockchain

Hybrid Blockchain ecosystem is similar to private Blockchain in terms of write access as well as maintain the consensus. In hybrid Blockchain, write access is restricted to certain nodes of the Blockchain network whereas consensus is basically maintained among a predefined group of nodes. Unlike private Blockchain, read access in hybrid Blockchain is usually open as in public Blockchain. Thus, Hybrid Blockchain provides the best of the other two types and more suitable for some specific applications such as stock exchange.

## III. EVOLUTION OF SECURITIES EXCHANGE

A securities exchange is a platform designed to enable the trading of shares of stock among traders with low transaction costs and high liquidity, as well as achieving other competitive advantages for its members. Mostly all securities trading takes place among the brokers or dealers. Securities exchanges are generally controlled both by pubic laws and regulations and by their own internal, rules. These norms lower the chance of fraud, default and mistake.

One country may have more than one securities exchange depending on geographical considerations as well as the needs of the financial market. The Amsterdam Stock Exchange was the first securities exchange in the world and was established in 1602. Examples of other leading stock exchanges include: New York Stock Exchange, NASDAQ, London Stock Exchange Group, Euronext, Hong Kong Stock Exchange, Japan Exchange Group and Shanghai Stock Exchange. At the end of 2017 the annual global stock market capitalization was $87.1 trillion, a 22.6% higher than as at end 2016 [23].

Because extremely high amount of transactions are being carried out in stock exchanges, it has always been a challenge to make the transactions secure and free of error. These high volumes of transactions in stock exchanges include order matching (e.g. in Hong Kong Stock Exchange about 1.24 billion transactions per day) and settlement transfers. Today, most trading is electronic, effected by remote data transfer, with records stored digitally. Thus, the security concerns regarding criminal hacking, as well as cyberwar, and political hacking (Hacktivism) have increased the need for raising the security level of the stock exchanges.

To explain the operational principles of securities exchanges, it is very important to look at the history, especially the evolution of securities exchanges and its reform brought by technologies. The evolution has moved from firm to market as private networks were replaced by public markets and now appears to be returning to firm again as private networks replace public markets. [20] The different points of historical development have been marked by different combinations of formal and informal institutions [21]. Although the shape of this evolution has been marked by the applicable law and the available technology, it has been considerably driven by the broker-dealer self-interest [22].

For about one thousand years, merchants of different size and kinds have traded securities amongst themselves. Thus the trades were performed directly. However, from 1800 to 2000, private clubs later made quasi-public "exchanges" took over an increasing portion of trading and actively monitored the trading process [22]. Around 2000, new information technology allowed electronic platforms to be introduced by the influential brokers, beginning the era of privately operated proprietary matching venues.

The founding principle of the first securities exchanges was to establish monopoly for their members. However, they also brought efficiency, and governments enacted laws to create transparency and established regulatory and monitoring systems. At the end of $20^{th}$ century, both new technology and regulatory reform let the largest broker-dealers escape the transparent egalitarianism earlier established by the exchanges and build their own proprietary trade matching platforms.

## IV. APPLICATION OF BLOCKCHAIN IN SECURITIES EXCHANGES

Leading securities exchanges are gradually embracing BC. NASDAQ has led the journey towards adoption of BC for stock exchanges [24]. ASX (Australian Securities Exchange) is also working towards replacing its current platform CHESS (Clearing House Electronic Subregister System) with BC by the end of 2020 or early 2021 for clearing, settlement and other post-trade services for Australian stocks [25]. In an attempt to cut cost, HKEX (Hong Kong Exchanges and Clearing) is seeking to implement Blockchain and now working with ASX to share their experience on BC implementation so far [26]. London Stock Exchange (LSE) is also working towards utilizing BC in a significant way. In July 2018, LSE has partnered with tech giant IBM which is considered as one of the global leaders in providing open-source blockchain solutions [24].

Based on the nature of securities exchange operations, we propose the use of a hybrid BC with Proof of Stack (PoS) for matching and randomised round robin approach for clearing and settlement. Matching of any buy and sell order of transaction can be carried by a central exchange function (which could be collectively maintained by broker-dealers) while clearing and settlement will still be controlled by the central counterparty clearing house (CCP). Thus the "back office" [1] tasks will be performed by the "closed circle" permitted nodes of CCP and will shift from highly centralized to round robin fashion.

Based on the technical anatomy of BC as presented in section II, BC if applied successfully in stock exchanges could bring real benefits.

- Since this hybrid approach of BC is distributed to some extent, this will provide better transparency consequences (for holders of the ledger) compared to the current approach being used in the securities exchanges. Although the level of transparency may not be as high as in public Blockchain.

- Blocks being chronologically added with the advent of cumulative hashing techniques as well as time-stamp, BC shall contribute towards a secure and trustworthy market.

- BC shall replace the exchange operator with the engineers building the BC system as intermediator. This shall establish trust amongst the traders and dealers to the extent the design and operation of the BC is fair, as transactions are verified and validated by the peers holding the ledger.

- Because BC technology could well be lower cost than legacy systems in place, and will require less maintenance, the transaction cost should eventually be lowered in the long run. Considering the investment required to replace an existing system with a new one, no cost savings will occur at the beginning. The cost of the new system will have to be paid for by users and customers for years, before they start getting the benefits.

- Settlement, in a securities exchange, takes place in cycles, intraday. In fact, most exchanges now have same-day settlement. The 2 or 3 day window in T+2 or T+3 [1] is to ensure that delivery of cash and shares is made. This is also to protect the interest of the brokers or dealers to "allow" multiple transfers of the same shares within a settlement cycle. On the contrary, China has instant settlement because proven cash must be available before trade is executed. BC can bring automation in the required post-trade activities. Securities can be settled in minutes instead of days i.e. real-time settlement with improved level of trust and transparency as well as supply chain optimisation and improved liquidity. However, same problem for settlement cycle may affect the overall process – are traders ready to deliver the securities and their purchase price?

- Considering the aforementioned accounting involved, because transaction costs will be lowered in the long run and post-trade inefficiencies could be reduced with the application of Blockchain, the market reorganization will likely attract new investment.

Due to the above mentioned benefits, BC seems to be attractive to both market participants and regulators. However, BC is still at its infancy and may be a source of regulatory and legal challenges which the regulators are still working to understand. BC may also give rise to legal and regulatory concerns about scalability and country specific data localisation requirements.

Another important regulatory issue is how trading, clearing and settlement are handled. In fact, they were considered separately in the past. However, Contemporary laws and regulations consider them to be one single transaction, but with many distinct steps while BC consolidates these elements together in one transaction.

Last but not least, law makers and regulatory bodies need to have a clear policy on "dematerialising" and "tracking claims" since they are perceived differently from a legal point of view. Transactions on DLT are technically tracking the claims of ownership transfers in dematerialised form, replacing a material object like a paper certificate by a digital entry. Digital tokens, which can also be transferred, could be legally classified as uncertificated securities for applying rules of legal transfer. However, until that is done, a token is not a share and has no legal bearing. This is because the token is an invention which falls outside the boundary of law relevant to share ownership. The owner of a share owns the share because it is registered in the owner's name on a legally approved share registry, which does not yet include the BC ledger or any similar DLT which facilitates tracking of digital tokens. Before a BC ledger or similar DLT can be considered as equivalent to legal share registry, the existing law will have to be changed issuing a relevant new statute.

## V. CONCLUDING DISCUSSIONS

"Pressure and budget to do something related to blockchains" [27] has been identified as one of the major reasons for implementing private Blockchain. Rather than following this trend, our approach was to step out of the crowd and investigate whether implementations of BC will make any real contributions to securities market operations. Considering the legal and technological aspects, to what extent can Blockchain technology add value to the securities settlement (registration) systems of securities exchanges as compared with the existing technologies? Our approach has been to investigate various variants of Blockchain technologies in detail to identify how Blockchain can help facilitate the transactions of a stock exchange. Based on the findings on BC technologies as well as considering the operational structure of securities exchanges, this paper puts forward a proposal for stock exchanges to adopt a hybrid approach of BC. Future research will involve exploring the legal aspects of such applications of Blockchain in Stock exchanges in more details.

ACKNOWLEDGEMENT *(Heading 5)*

This research is funded by the Centre for Financial Regulation and Economic Development (CFRED), The Chinese University of Hong Kong (CUHK), Hong Kong.